\documentclass[aps,floatfix,prl,showpacs,notitlepage,reprint]{revtex4-1}
\usepackage{graphicx}
\usepackage{dcolumn}
\usepackage{amsmath}
\usepackage{bm}
\usepackage[utf8]{inputenc}
\usepackage{amsmath}
\usepackage{natbib}
\usepackage{graphicx}

\newcommand{\beq}{\begin{equation}}
\newcommand{\eeq}{\end {equation}}
\newcommand{\bea}{\begin{eqnarray}}
\newcommand{\eea}{\end{eqnarray}}

\begin{document}
\title{Photo-Induced Image Current}

\author{K. Koksal and F.A. Celik} 

\affiliation{Physics Department, Bitlis Eren University, Bitlis 13000, Turkey}
\date{\today}

\begin{abstract}
We study the possibility of the generation of the photo-induced image currents at a distance from the surface of nano-sized metal clusters by using time-dependent perturbation theory. We reveal that the wave function of an electron excited to the image state is localized outside the surface and current flows in a spherical shell whose radius is a few times the radius of the sphere. Spin polarized light has been applied to a perfect icosahedral metal cluster Li$_{13}$ whose optimization is achieved by molecular dynamic simulation and band structure is obtained by DFT method and by solution of radial Schrödinger equation. Up to our knowledge, despite the great effort on their characteristics, image electrons have not been the subject of the studies on photo-induced current.  
\end{abstract}


\maketitle
The optical properties of small metal clusters have been intensively studied due to the significance for nanooptics\cite{ghosh2007interparticle} and medical diagnostics \cite{unser2015localized}. And they are also currently used in nanoscale sensing applications \cite{liu2020one} because of the sensitivity of their localized surface plasmon properties to the size, shape or environment. A metal surface is the reason of the forming an infinite number of Rydberg-type energy levels near the vacuum level namely image (potential) states \cite{pendry1980new}. The surface boundary like a mirror allows the formation of a positive image charge inside the cluster to balance the electron excited to the image state and residing outside the surface  \cite{weinert1985image}. The theoretical and experimental studies on low-dimensional metal and molecular structures reveal the existence and striking properties of image states for nanotubes \cite{granger2002highly, zamkov2004time}, nanowire lattices    \cite{segal2005tunable}, quantum dots \cite{craes2013mapping} and $2$D layers \cite{silkin2009image,liu2021nanoscale}. 
\begin{figure}
  \centering
  \includegraphics[width=.6\columnwidth]{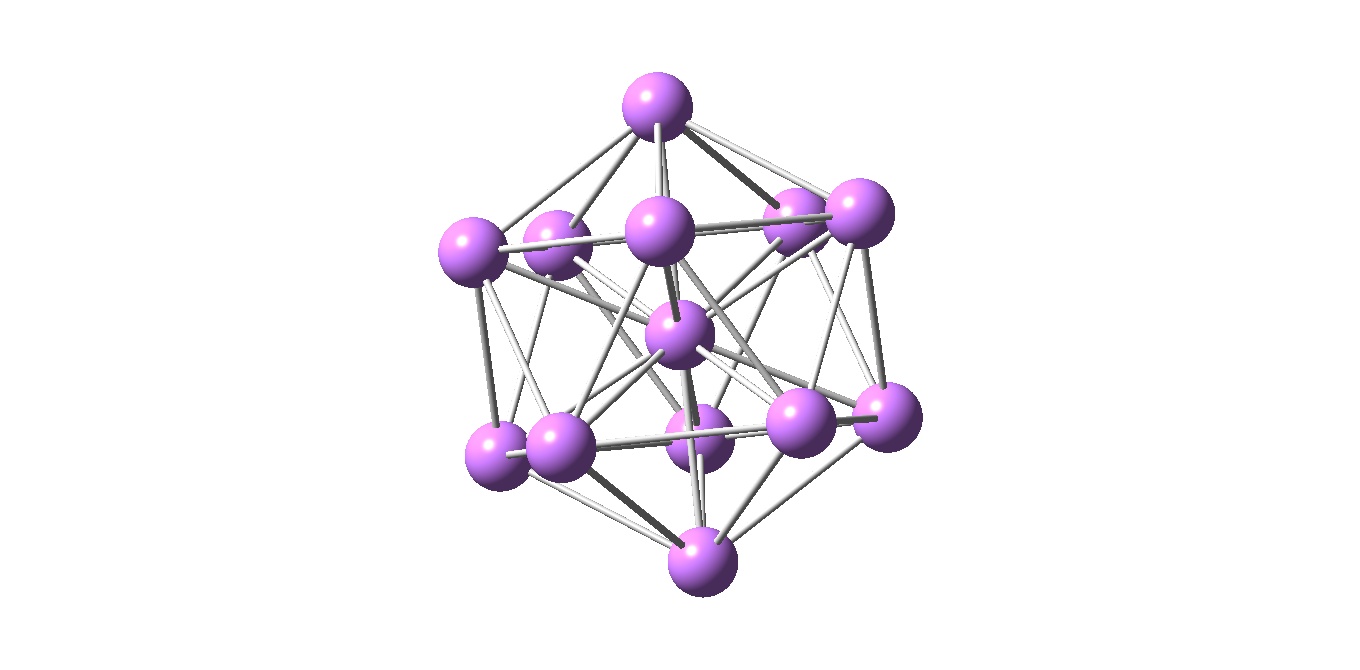}  
  \caption{Icosahedral Li$_{13}$ cluster obtained from the molecular dynamic simulation. }
  \label{Li13}
\end{figure}

This work is inspired from the exciting results of some works on image states such that: image states have spin properties \cite{tognolini2015rashba,nekovee1993magnetic,winkelmann2007ultrafast}, there is the possibility of  high amount of image electron spin polarization \cite{winkelmann2007ultrafast}, an image electron is able to propagate into the metal cluster without collision \cite{winter2011trapping}, it has long lifetime in low dimensional structures \cite{granger2002highly}.  In the light of these motivating works, we aim to investigate the image potential states, the possibility of optical spin transfer and the production of the photo-induced persistent ring current for a simple, spherical, small metal cluster. We focus to observe a ring current around the cluster which flow far from outside the surface. One way of the induction of ring current is using optical angular momentum transfer to the ring-shaped structures \cite{barth2006unidirectional,koksal2016optical,koksal2017effect,koksal2017spin,kocc2015quantum,kocc2017mapping}.
\begin{figure}
  \centering
  \includegraphics[width=1\columnwidth]{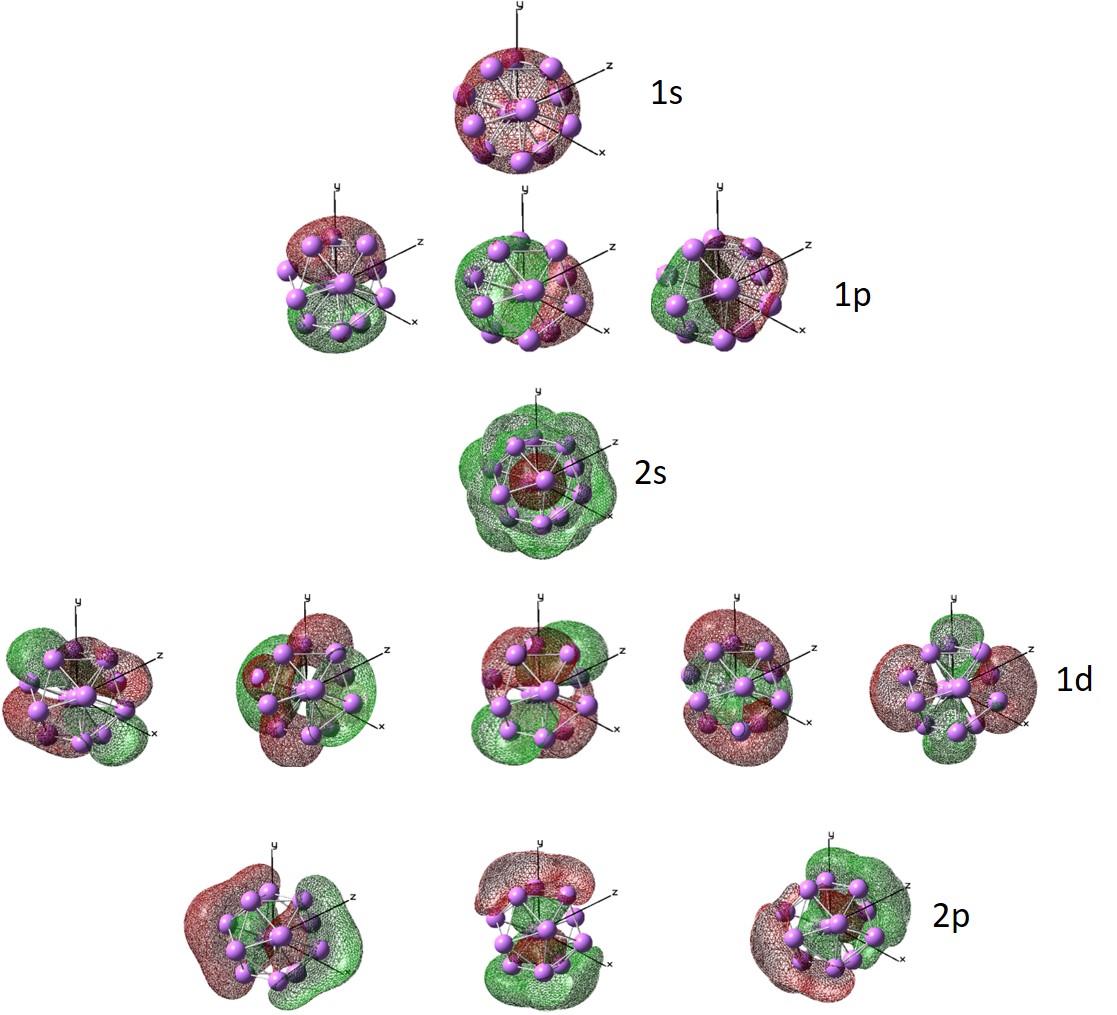}  
  \caption{The molecular orbital wavefunctions of  Li$_{13}$ cluster obtained from the molecular dynamic simulation.}
  \label{Li13wf}
\end{figure}

In this work, we aim to analyze the production of image currents arising from the image electronic states which are localized outside the surface. According to the theory, an applied spin-polarized electromagnetic field leads to an excitation of the bulk electrons to the unoccupied surface or image states. If the electron is excited to the level having a net magnetic quantum number, it causes an induced current which is flowing outside the surface. 

\begin{figure}
  \centering
  \includegraphics[width=1\columnwidth]{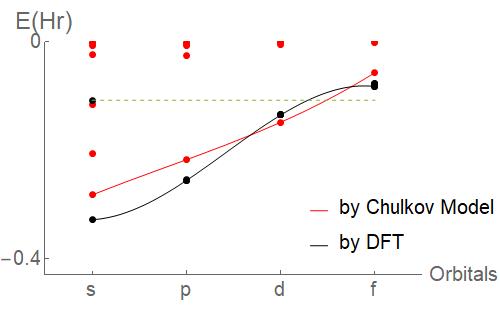}  
  \caption{The comparison of energy eigenvalues obtained two different models.}
  \label{band}
\end{figure}

In our theoretical calculations, as a first step, by performing temperature dependent molecular dynamic simulation of $5000$ Li atoms, we obtained a crystal of Li and choose the most appropriate icosahedral clusters in the crystal. We transferred the structure properties of this cluster to the Gaussian $09$  and we optimized again. We showed that the optimization has not changed the structure property of the cluster. We used the DFT technique in the frame of B3LYP scheme to obtain the electronic structure of the cluster. Then we solved the Schrödinger's equation to get the image states. The Fig. \ref{Li13wf} shows the molecular orbitals for different states of a metal cluster which has a perfect spherical symmetry. In order to obtain the radial profile and energy levels of surface and image states we solve the radial Schrödinger equation with the empirical potential introduced by Chulkov et. al. \cite{chulkov1997image,chulkov1999image,silkin2009image}. This model potential is the result of ab-initio
techniques and experimental observations, which can be
written as  
\begin{equation}
\begin{split}
V_1(r)&=A_{10}+A_1 \cos{\frac{2\pi}{a_s}r},\;\; r<0\\
V_2(r)&=-A_{20}+A_2 \cos{\beta r},\;\; 0<r<r_{im}\\
V_3(r)&=A_3 e^{-\alpha(r-r_1)},\;\; r_1<r<r_{im}\\
V_4(r)&=\frac{\exp[-\lambda(r-r_{im})]-1}{4(r-r_{im})},\;\; r_{im}<z
\label{Eq:potential}
\end{split}
\end{equation}%
This one-electron potential is a result of local density approximation. For $z>z_{im}$, image potential has been used instead of the LDA potential \cite{chulkov1999image}.

What we expect from the electronic band structure of a spherical cluster is to provide a consistency between the bulk levels obtained from DFT and solution of radial Schrödinger equation. The comparison of energy eigenvalues of the states obtained from two different methods has been shown in the Fig. \ref{band}. This figure indicates the consistency of the results obtained different methods. The wavefunctions obtained from Chulkov model will be used in image current density calculations.

\begin{figure}
  \centering
  \includegraphics[width=1\columnwidth]{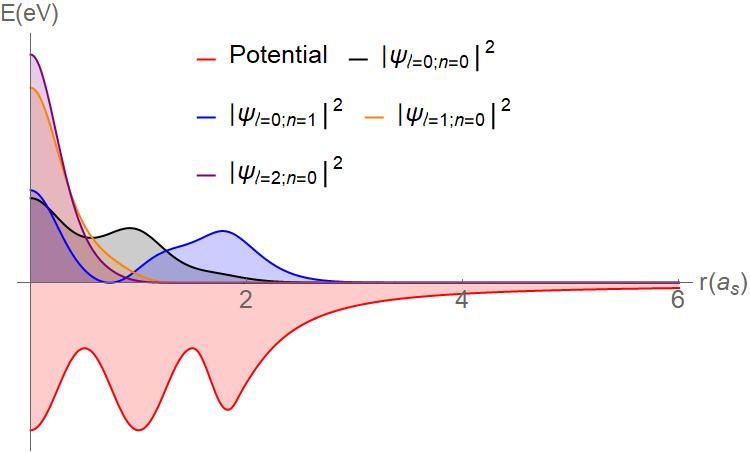}  
  \caption{The radial potential and corresponding radial wave functions of bulk states of lithium cluster. These wavefunctions belong to the occupied electrons which are localized in the potential region.}
  \label{Li13wf1}
\end{figure}
From the Fig. \ref{band}, we can see the possible excitation between the bulk and surface or image states. In the case of circularly polarized light, $p-s$ or $p-d$ excitations are possible. An electron excited to the $s$ state is not able to induce a ring current because $s$ state does not include magnetic quantum number as will be seen Eq. \ref{Eq:curden2}. An excitation of an electron to $d$ state has very low probability because of small overlapping function between $p$ and $d$ states. As a result of our calculations, we found that the highest overlapping value is obtained in $s-p$ transitions. Therefore, we decided to calculate the ring current induced by $1s-2p$ and $1s-3p$ transitions. Here, $1s$, $2p$ and $3p$ refer to the bulk, surface and image states.       

\begin{figure}
  \centering
  \includegraphics[width=1\columnwidth]{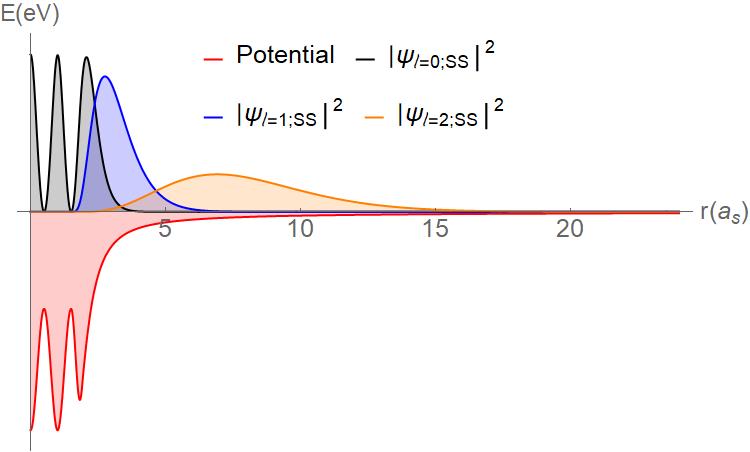}  
  \caption{The surface states for different $\ell$ values. In the case of $\ell=0$, the surface state wavefunction of this spherical cluster is  very similar to that of flat surface. For different values of $\ell$, the surface states are pushed through the outer region of the surface by barrier potential. }
  \label{Li13wf2}
\end{figure}
\begin{figure}
  \centering
  \includegraphics[width=1\columnwidth]{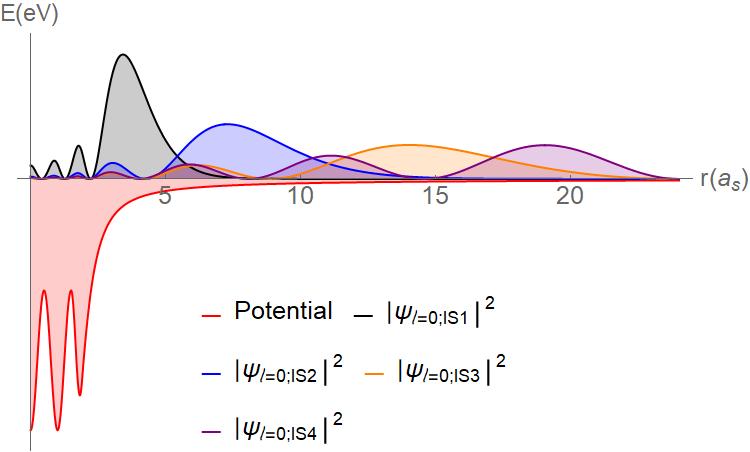}  
  \caption{The image states of the spherical cluster for $\ell=0$. Here $4.$ image state is mostly localized in a distance which is almost $14$ times more than the radius of the cluster. In this case only believable value is first image state.}
  \label{Li13wf3}
\end{figure}
\begin{figure}
  \centering
  \includegraphics[width=1\columnwidth]{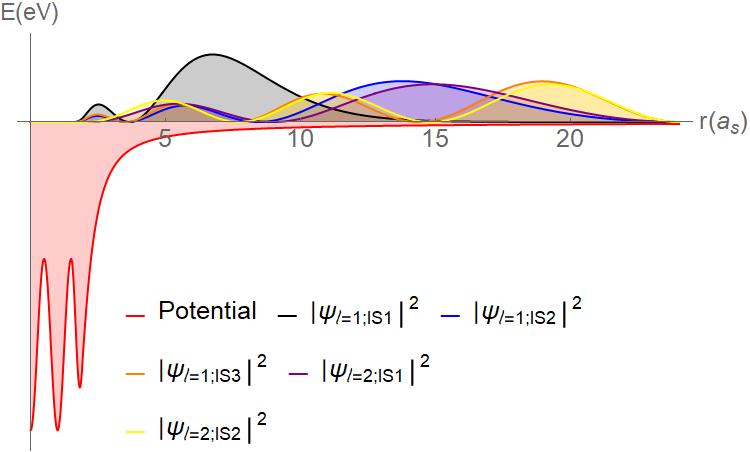}  
  \caption{The image states of the spherical cluster for $\ell>0$. We only use the image state $\psi_{|\ell=1;IS=1|}$}
  \label{Li13wf4}
\end{figure}
In order to analyze the characteristics of the occupied and unoccupied states, we will give some attention to the figures which describe the bulk, surface and image states. Fig. \ref{Li13wf1} show the wavefunction of electrons in the occupied states. $l=0$, $l=1$ and $l=2$ refer to the $s$, $p$ and $d$ states. The radial distribution of the electron wavefunction falls within the radial potential region. 

Fig. \ref{Li13wf2} indicates the surface electron states which are unoccupied. Because the surface and image states are broad in position space, the scale needs to be expanded as in the figure. The profile of the surface state for $l=0$ is very similar to that of the flat surface. For $l=1$, the surface state wavefunction is mostly localized out of the surface. And the position of the peak of this state is at $r=3.5\;a_s$. For $l=2$, the surface state wavefunction is localized far outside the surface. The peak position of $l=2$ wavefunction is almost $7\;a_s$. 

Fig. \ref{Li13wf3} shows the wavefunctions of image states for $l=0$. The closer to the vacuum level, the further the localization moves away from the surface. Albeit fantastic is an electron traveling almost $10$ times the diameter of the sphere from the surface of a small metal sphere, a similar investigation \cite{granger2002highly} has reported that the image wavefunctions are localized at a distance which is $5$ times the diameter of a carbon nanotube. Fig. \ref{Li13wf4} shows other image state wavefunctions. In our calculations we only used the first image state for $l=1$.

In the study of Koksal et. al. \cite{koksal2012charge}, the ring currents have been calculated by using time-dependent perturbation theory. Referring the details to this work, we only mention the last equation which is
\begin{equation}
\begin{split}
&\mathbf{j_{\phi}}(\mathbf{r})=\mathbf{\Tilde{j}}_{\phi}\times \mathbf{M}_{n l m,n_0 l_0 m_0}\\
&\mathbf{\Tilde{j}}_{\phi}=\frac{m \hbar}{m_e} \frac{\left|\psi_{n l m}(r,\theta,\phi)\right|^2}{r \sin{\theta}}  \mathbf{\hat{\phi}} \\
&\mathbf{M}_{n l m,n_0 l_0 m_0}=\frac{4}{\delta^2} A_0^2 (E_{n l m}-E_{n_0 l_0 m_0})^2 \times\\
& \left|\left<n l m \;|\;  r \cos\theta (\cos\phi+i \sin\phi)     \;|\;n_0 l_0 m_0\right>\right|^2
\label{Eq:curden2}
\end{split}
\end{equation}
where $\mathbf{\Tilde{j}}_{\phi}$ corresponds to the current density due to the excited electron. Other term $\mathbf{M}_{n l m,n_0 l_0 m_0}$ shows the transition probability. $A_0$ is the amplitude of the electric field, $E_{n l m}$ is energy of the excited states. $n$,$l$,$m$ are radial, orbital and magnetic quantum numbers, respectively. $\delta$ refers transition rate, $m_e$ and $\hbar$ are electron mass and Planck's constant, respectively. The expression in the parenthesis $(\cos\phi+i \sin\phi)$ refers to the spin polarization of the light.  

In our calculations we only consider one electron transition between ground state into the image-potential state or surface state. The frequency of the light is tuned to be the same as the difference between the energies of intended bulk and surface/image states.   Polarizarion vector is $\mathbf{\hat{e}}=(\cos \phi,i \sin \phi, 0)$.

\begin{figure}
  \centering
  \includegraphics[width=1\columnwidth]{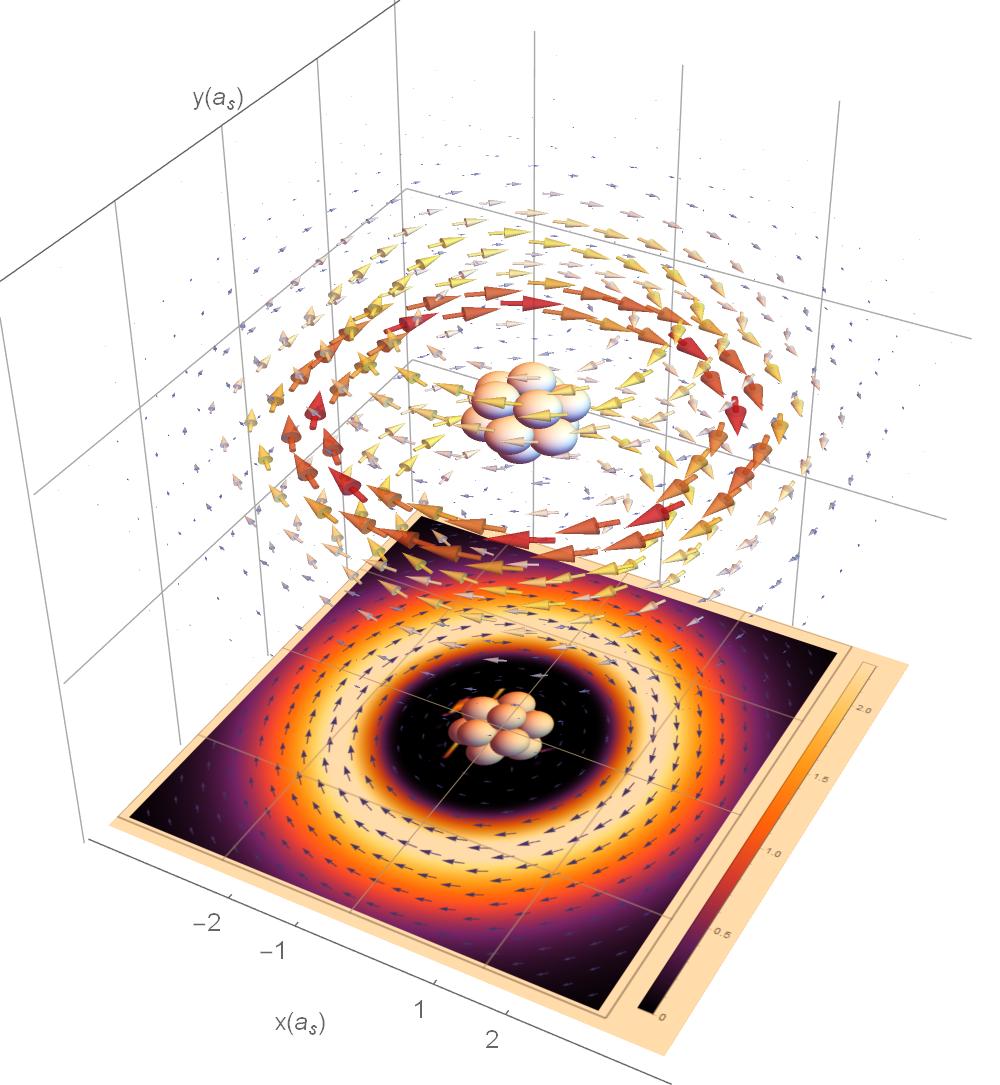}  
  \caption{The photo-induced current is due to the transition of the bulk electron in the  bulk state $\psi_{l=0;n=1}$ into the surface  state $\psi_{l=1;SS}$ (please look at the Fig. \ref{Li13wf2}). The unit of the current is in the range of $\mu$A. The energy value of bulk and surface states are $-7.6859$ Hr and $-0.7141$ Hr, respectively.}
  \label{imagecurrentSS}
\end{figure}
According to the scenario,  spin polarized beam is applied to the Li$_{13}$ spherical cluster to manage the transfer of the photon spin into the electron in bulk state. In order to observe a net ring current, the electron should be kicked to an electronic level which has orbital $l\neq 0$ and magnetic $m\neq 0$ quantum numbers. An $s-p$ transition will be sufficient to observe a ring current and a circulation of an electron. When we manage to excite the electron surface or image state which are localized outside the surface, it is possible to observe a ring current in the region out of the surface.

\begin{figure}
  \centering
  \includegraphics[width=1\columnwidth]{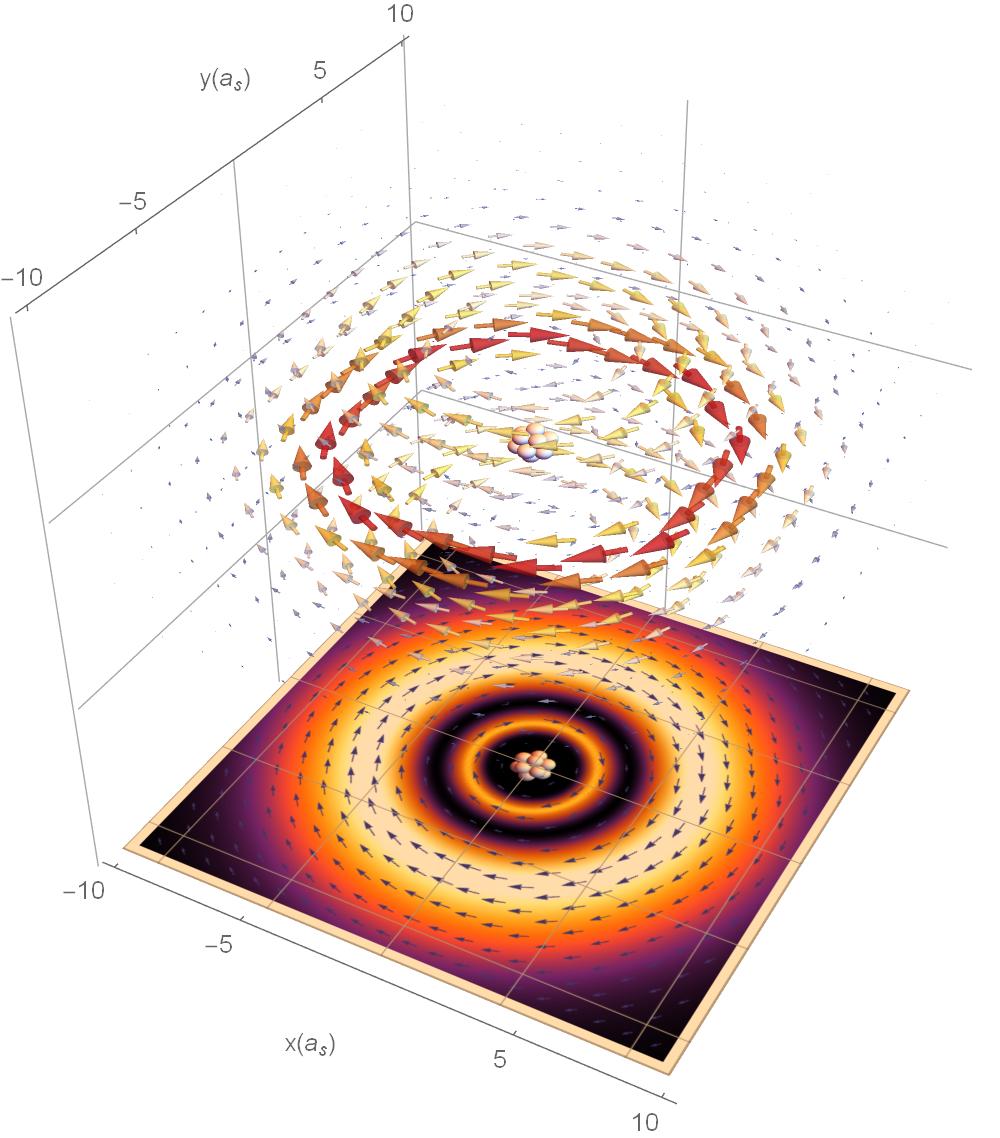}  
  \caption{This current is due to the transition of an electron between bulk state $\psi_{l=0;n=1}$ and image potential state ${\psi_{l=1;IS1}}$ (please look at the Fig. \ref{Li13wf4}). The energy value of bulk and image states are $-7.6859$ Hr and $-0.1829$ Hr, respectively. $1 a.u.$ for the current means $6.6$ mA. The current here is almost $10$ $\mu$A flowing in clockwise direction. The maximum value of current density is 2.35 $\mu$A/nm$^2$. }
  \label{imagecurrentIS}
\end{figure}

As seen from the Figs. \ref{imagecurrentSS} and \ref{imagecurrentIS}, the photo-induced image(surface) electron current density is distributed symmetrically outside the surface of the metal cluster. This current is the result of the excitation of an electron in bulk state into the surface or image state, $\psi_{l=1}$. Although overlap function between a bulk state and an image/surface state is small, the amount of obtained induced surface current is around $\mu$A, that of induced image current is around nA. 

The properties of the image states are similar to those of the hydrogenic wavefunctions and there exists infinite Rydeberg-type series. According to this analogy and as a result of our calculations, a photo-induced image electron current refers to the electron rotating around a nucleus of which the constituent is Li$_{13}$ cluster instead of proton. We believe that this theoretical work can be conformed by well designed experimental studies. 







\bibliography{references} 

\begin{thebibliography}{24}%
\makeatletter
\providecommand \@ifxundefined [1]{%
 \@ifx{#1\undefined}
}%
\providecommand \@ifnum [1]{%
 \ifnum #1\expandafter \@firstoftwo
 \else \expandafter \@secondoftwo
 \fi
}%
\providecommand \@ifx [1]{%
 \ifx #1\expandafter \@firstoftwo
 \else \expandafter \@secondoftwo
 \fi
}%
\providecommand \natexlab [1]{#1}%
\providecommand \enquote  [1]{``#1''}%
\providecommand \bibnamefont  [1]{#1}%
\providecommand \bibfnamefont [1]{#1}%
\providecommand \citenamefont [1]{#1}%
\providecommand \href@noop [0]{\@secondoftwo}%
\providecommand \href [0]{\begingroup \@sanitize@url \@href}%
\providecommand \@href[1]{\@@startlink{#1}\@@href}%
\providecommand \@@href[1]{\endgroup#1\@@endlink}%
\providecommand \@sanitize@url [0]{\catcode `\\12\catcode `\$12\catcode
  `\&12\catcode `\#12\catcode `\^12\catcode `\_12\catcode `\%12\relax}%
\providecommand \@@startlink[1]{}%
\providecommand \@@endlink[0]{}%
\providecommand \url  [0]{\begingroup\@sanitize@url \@url }%
\providecommand \@url [1]{\endgroup\@href {#1}{\urlprefix }}%
\providecommand \urlprefix  [0]{URL }%
\providecommand \Eprint [0]{\href }%
\providecommand \doibase [0]{http://dx.doi.org/}%
\providecommand \selectlanguage [0]{\@gobble}%
\providecommand \bibinfo  [0]{\@secondoftwo}%
\providecommand \bibfield  [0]{\@secondoftwo}%
\providecommand \translation [1]{[#1]}%
\providecommand \BibitemOpen [0]{}%
\providecommand \bibitemStop [0]{}%
\providecommand \bibitemNoStop [0]{.\EOS\space}%
\providecommand \EOS [0]{\spacefactor3000\relax}%
\providecommand \BibitemShut  [1]{\csname bibitem#1\endcsname}%
\let\auto@bib@innerbib\@empty
\bibitem [{\citenamefont {Ghosh}\ and\ \citenamefont
  {Pal}(2007)}]{ghosh2007interparticle}%
  \BibitemOpen
  \bibfield  {author} {\bibinfo {author} {\bibfnamefont {S.~K.}\ \bibnamefont
  {Ghosh}}\ and\ \bibinfo {author} {\bibfnamefont {T.}~\bibnamefont {Pal}},\
  }\href@noop {} {\bibfield  {journal} {\bibinfo  {journal} {Chemical reviews}\
  }\textbf {\bibinfo {volume} {107}},\ \bibinfo {pages} {4797} (\bibinfo {year}
  {2007})}\BibitemShut {NoStop}%
\bibitem [{\citenamefont {Unser}\ \emph {et~al.}(2015)\citenamefont {Unser},
  \citenamefont {Bruzas}, \citenamefont {He},\ and\ \citenamefont
  {Sagle}}]{unser2015localized}%
  \BibitemOpen
  \bibfield  {author} {\bibinfo {author} {\bibfnamefont {S.}~\bibnamefont
  {Unser}}, \bibinfo {author} {\bibfnamefont {I.}~\bibnamefont {Bruzas}},
  \bibinfo {author} {\bibfnamefont {J.}~\bibnamefont {He}}, \ and\ \bibinfo
  {author} {\bibfnamefont {L.}~\bibnamefont {Sagle}},\ }\href@noop {}
  {\bibfield  {journal} {\bibinfo  {journal} {Sensors}\ }\textbf {\bibinfo
  {volume} {15}},\ \bibinfo {pages} {15684} (\bibinfo {year}
  {2015})}\BibitemShut {NoStop}%
\bibitem [{\citenamefont {Liu}\ and\ \citenamefont {Ma}(2020)}]{liu2020one}%
  \BibitemOpen
  \bibfield  {author} {\bibinfo {author} {\bibfnamefont {Y.}~\bibnamefont
  {Liu}}\ and\ \bibinfo {author} {\bibfnamefont {Y.}~\bibnamefont {Ma}},\
  }\href@noop {} {\bibfield  {journal} {\bibinfo  {journal} {Frontiers in
  Physics}\ }\textbf {\bibinfo {volume} {8}},\ \bibinfo {pages} {312} (\bibinfo
  {year} {2020})}\BibitemShut {NoStop}%
\bibitem [{\citenamefont {Pendry}(1980)}]{pendry1980new}%
  \BibitemOpen
  \bibfield  {author} {\bibinfo {author} {\bibfnamefont {J.}~\bibnamefont
  {Pendry}},\ }\href@noop {} {\bibfield  {journal} {\bibinfo  {journal}
  {Physical Review Letters}\ }\textbf {\bibinfo {volume} {45}},\ \bibinfo
  {pages} {1356} (\bibinfo {year} {1980})}\BibitemShut {NoStop}%
\bibitem [{\citenamefont {Weinert}\ \emph {et~al.}(1985)\citenamefont
  {Weinert}, \citenamefont {Hulbert},\ and\ \citenamefont
  {Johnson}}]{weinert1985image}%
  \BibitemOpen
  \bibfield  {author} {\bibinfo {author} {\bibfnamefont {M.}~\bibnamefont
  {Weinert}}, \bibinfo {author} {\bibfnamefont {S.}~\bibnamefont {Hulbert}}, \
  and\ \bibinfo {author} {\bibfnamefont {P.}~\bibnamefont {Johnson}},\
  }\href@noop {} {\bibfield  {journal} {\bibinfo  {journal} {Physical review
  letters}\ }\textbf {\bibinfo {volume} {55}},\ \bibinfo {pages} {2055}
  (\bibinfo {year} {1985})}\BibitemShut {NoStop}%
\bibitem [{\citenamefont {Granger}\ \emph {et~al.}(2002)\citenamefont
  {Granger}, \citenamefont {Kr{\'a}l}, \citenamefont {Sadeghpour},\ and\
  \citenamefont {Shapiro}}]{granger2002highly}%
  \BibitemOpen
  \bibfield  {author} {\bibinfo {author} {\bibfnamefont {B.~E.}\ \bibnamefont
  {Granger}}, \bibinfo {author} {\bibfnamefont {P.}~\bibnamefont {Kr{\'a}l}},
  \bibinfo {author} {\bibfnamefont {H.}~\bibnamefont {Sadeghpour}}, \ and\
  \bibinfo {author} {\bibfnamefont {M.}~\bibnamefont {Shapiro}},\ }\href@noop
  {} {\bibfield  {journal} {\bibinfo  {journal} {Physical review letters}\
  }\textbf {\bibinfo {volume} {89}},\ \bibinfo {pages} {135506} (\bibinfo
  {year} {2002})}\BibitemShut {NoStop}%
\bibitem [{\citenamefont {Zamkov}\ \emph {et~al.}(2004)\citenamefont {Zamkov},
  \citenamefont {Woody}, \citenamefont {Bing}, \citenamefont {Chakraborty},
  \citenamefont {Chang}, \citenamefont {Thumm},\ and\ \citenamefont
  {Richard}}]{zamkov2004time}%
  \BibitemOpen
  \bibfield  {author} {\bibinfo {author} {\bibfnamefont {M.}~\bibnamefont
  {Zamkov}}, \bibinfo {author} {\bibfnamefont {N.}~\bibnamefont {Woody}},
  \bibinfo {author} {\bibfnamefont {S.}~\bibnamefont {Bing}}, \bibinfo {author}
  {\bibfnamefont {H.}~\bibnamefont {Chakraborty}}, \bibinfo {author}
  {\bibfnamefont {Z.}~\bibnamefont {Chang}}, \bibinfo {author} {\bibfnamefont
  {U.}~\bibnamefont {Thumm}}, \ and\ \bibinfo {author} {\bibfnamefont
  {P.}~\bibnamefont {Richard}},\ }\href@noop {} {\bibfield  {journal} {\bibinfo
   {journal} {Physical review letters}\ }\textbf {\bibinfo {volume} {93}},\
  \bibinfo {pages} {156803} (\bibinfo {year} {2004})}\BibitemShut {NoStop}%
\bibitem [{\citenamefont {Segal}\ \emph {et~al.}(2005)\citenamefont {Segal},
  \citenamefont {Granger}, \citenamefont {Sadeghpour}, \citenamefont
  {Kr{\'a}l},\ and\ \citenamefont {Shapiro}}]{segal2005tunable}%
  \BibitemOpen
  \bibfield  {author} {\bibinfo {author} {\bibfnamefont {D.}~\bibnamefont
  {Segal}}, \bibinfo {author} {\bibfnamefont {B.~E.}\ \bibnamefont {Granger}},
  \bibinfo {author} {\bibfnamefont {H.}~\bibnamefont {Sadeghpour}}, \bibinfo
  {author} {\bibfnamefont {P.}~\bibnamefont {Kr{\'a}l}}, \ and\ \bibinfo
  {author} {\bibfnamefont {M.}~\bibnamefont {Shapiro}},\ }\href@noop {}
  {\bibfield  {journal} {\bibinfo  {journal} {Physical review letters}\
  }\textbf {\bibinfo {volume} {94}},\ \bibinfo {pages} {016402} (\bibinfo
  {year} {2005})}\BibitemShut {NoStop}%
\bibitem [{\citenamefont {Craes}\ \emph {et~al.}(2013)\citenamefont {Craes},
  \citenamefont {Runte}, \citenamefont {Klinkhammer}, \citenamefont {Kralj},
  \citenamefont {Michely},\ and\ \citenamefont {Busse}}]{craes2013mapping}%
  \BibitemOpen
  \bibfield  {author} {\bibinfo {author} {\bibfnamefont {F.}~\bibnamefont
  {Craes}}, \bibinfo {author} {\bibfnamefont {S.}~\bibnamefont {Runte}},
  \bibinfo {author} {\bibfnamefont {J.}~\bibnamefont {Klinkhammer}}, \bibinfo
  {author} {\bibfnamefont {M.}~\bibnamefont {Kralj}}, \bibinfo {author}
  {\bibfnamefont {T.}~\bibnamefont {Michely}}, \ and\ \bibinfo {author}
  {\bibfnamefont {C.}~\bibnamefont {Busse}},\ }\href@noop {} {\bibfield
  {journal} {\bibinfo  {journal} {Physical review letters}\ }\textbf {\bibinfo
  {volume} {111}},\ \bibinfo {pages} {056804} (\bibinfo {year}
  {2013})}\BibitemShut {NoStop}%
\bibitem [{\citenamefont {Silkin}\ \emph {et~al.}(2009)\citenamefont {Silkin},
  \citenamefont {Zhao}, \citenamefont {Guinea}, \citenamefont {Chulkov},
  \citenamefont {Echenique},\ and\ \citenamefont {Petek}}]{silkin2009image}%
  \BibitemOpen
  \bibfield  {author} {\bibinfo {author} {\bibfnamefont {V.}~\bibnamefont
  {Silkin}}, \bibinfo {author} {\bibfnamefont {J.}~\bibnamefont {Zhao}},
  \bibinfo {author} {\bibfnamefont {F.}~\bibnamefont {Guinea}}, \bibinfo
  {author} {\bibfnamefont {E.}~\bibnamefont {Chulkov}}, \bibinfo {author}
  {\bibfnamefont {P.}~\bibnamefont {Echenique}}, \ and\ \bibinfo {author}
  {\bibfnamefont {H.}~\bibnamefont {Petek}},\ }\href@noop {} {\bibfield
  {journal} {\bibinfo  {journal} {Physical Review B}\ }\textbf {\bibinfo
  {volume} {80}},\ \bibinfo {pages} {121408} (\bibinfo {year}
  {2009})}\BibitemShut {NoStop}%
\bibitem [{\citenamefont {Liu}\ \emph {et~al.}(2021)\citenamefont {Liu},
  \citenamefont {Wang}, \citenamefont {Yakobson},\ and\ \citenamefont
  {Hersam}}]{liu2021nanoscale}%
  \BibitemOpen
  \bibfield  {author} {\bibinfo {author} {\bibfnamefont {X.}~\bibnamefont
  {Liu}}, \bibinfo {author} {\bibfnamefont {L.}~\bibnamefont {Wang}}, \bibinfo
  {author} {\bibfnamefont {B.~I.}\ \bibnamefont {Yakobson}}, \ and\ \bibinfo
  {author} {\bibfnamefont {M.~C.}\ \bibnamefont {Hersam}},\ }\href@noop {}
  {\bibfield  {journal} {\bibinfo  {journal} {Nano letters}\ }\textbf {\bibinfo
  {volume} {21}},\ \bibinfo {pages} {1169} (\bibinfo {year}
  {2021})}\BibitemShut {NoStop}%
\bibitem [{\citenamefont {Tognolini}\ \emph {et~al.}(2015)\citenamefont
  {Tognolini}, \citenamefont {Achilli}, \citenamefont {Longetti}, \citenamefont
  {Fava}, \citenamefont {Mariani}, \citenamefont {Trioni},\ and\ \citenamefont
  {Pagliara}}]{tognolini2015rashba}%
  \BibitemOpen
  \bibfield  {author} {\bibinfo {author} {\bibfnamefont {S.}~\bibnamefont
  {Tognolini}}, \bibinfo {author} {\bibfnamefont {S.}~\bibnamefont {Achilli}},
  \bibinfo {author} {\bibfnamefont {L.}~\bibnamefont {Longetti}}, \bibinfo
  {author} {\bibfnamefont {E.}~\bibnamefont {Fava}}, \bibinfo {author}
  {\bibfnamefont {C.}~\bibnamefont {Mariani}}, \bibinfo {author} {\bibfnamefont
  {M.}~\bibnamefont {Trioni}}, \ and\ \bibinfo {author} {\bibfnamefont
  {S.}~\bibnamefont {Pagliara}},\ }\href@noop {} {\bibfield  {journal}
  {\bibinfo  {journal} {Physical review letters}\ }\textbf {\bibinfo {volume}
  {115}},\ \bibinfo {pages} {046801} (\bibinfo {year} {2015})}\BibitemShut
  {NoStop}%
\bibitem [{\citenamefont {Nekovee}\ \emph {et~al.}(1993)\citenamefont
  {Nekovee}, \citenamefont {Crampin},\ and\ \citenamefont
  {Inglesfield}}]{nekovee1993magnetic}%
  \BibitemOpen
  \bibfield  {author} {\bibinfo {author} {\bibfnamefont {M.}~\bibnamefont
  {Nekovee}}, \bibinfo {author} {\bibfnamefont {S.}~\bibnamefont {Crampin}}, \
  and\ \bibinfo {author} {\bibfnamefont {J.}~\bibnamefont {Inglesfield}},\
  }\href@noop {} {\bibfield  {journal} {\bibinfo  {journal} {Physical review
  letters}\ }\textbf {\bibinfo {volume} {70}},\ \bibinfo {pages} {3099}
  (\bibinfo {year} {1993})}\BibitemShut {NoStop}%
\bibitem [{\citenamefont {Winkelmann}\ \emph {et~al.}(2007)\citenamefont
  {Winkelmann}, \citenamefont {Bisio}, \citenamefont {Ocana}, \citenamefont
  {Lin}, \citenamefont {N{\`y}vlt}, \citenamefont {Petek},\ and\ \citenamefont
  {Kirschner}}]{winkelmann2007ultrafast}%
  \BibitemOpen
  \bibfield  {author} {\bibinfo {author} {\bibfnamefont {A.}~\bibnamefont
  {Winkelmann}}, \bibinfo {author} {\bibfnamefont {F.}~\bibnamefont {Bisio}},
  \bibinfo {author} {\bibfnamefont {R.}~\bibnamefont {Ocana}}, \bibinfo
  {author} {\bibfnamefont {W.-C.}\ \bibnamefont {Lin}}, \bibinfo {author}
  {\bibfnamefont {M.}~\bibnamefont {N{\`y}vlt}}, \bibinfo {author}
  {\bibfnamefont {H.}~\bibnamefont {Petek}}, \ and\ \bibinfo {author}
  {\bibfnamefont {J.}~\bibnamefont {Kirschner}},\ }\href@noop {} {\bibfield
  {journal} {\bibinfo  {journal} {Physical review letters}\ }\textbf {\bibinfo
  {volume} {98}},\ \bibinfo {pages} {226601} (\bibinfo {year}
  {2007})}\BibitemShut {NoStop}%
\bibitem [{\citenamefont {Winter}\ \emph {et~al.}(2011)\citenamefont {Winter},
  \citenamefont {Chulkov},\ and\ \citenamefont
  {H{\"o}fer}}]{winter2011trapping}%
  \BibitemOpen
  \bibfield  {author} {\bibinfo {author} {\bibfnamefont {M.}~\bibnamefont
  {Winter}}, \bibinfo {author} {\bibfnamefont {E.~V.}\ \bibnamefont {Chulkov}},
  \ and\ \bibinfo {author} {\bibfnamefont {U.}~\bibnamefont {H{\"o}fer}},\
  }\href@noop {} {\bibfield  {journal} {\bibinfo  {journal} {Physical review
  letters}\ }\textbf {\bibinfo {volume} {107}},\ \bibinfo {pages} {236801}
  (\bibinfo {year} {2011})}\BibitemShut {NoStop}%
\bibitem [{\citenamefont {Barth}\ \emph {et~al.}(2006)\citenamefont {Barth},
  \citenamefont {Manz}, \citenamefont {Shigeta},\ and\ \citenamefont
  {Yagi}}]{barth2006unidirectional}%
  \BibitemOpen
  \bibfield  {author} {\bibinfo {author} {\bibfnamefont {I.}~\bibnamefont
  {Barth}}, \bibinfo {author} {\bibfnamefont {J.}~\bibnamefont {Manz}},
  \bibinfo {author} {\bibfnamefont {Y.}~\bibnamefont {Shigeta}}, \ and\
  \bibinfo {author} {\bibfnamefont {K.}~\bibnamefont {Yagi}},\ }\href@noop {}
  {\bibfield  {journal} {\bibinfo  {journal} {Journal of the American Chemical
  Society}\ }\textbf {\bibinfo {volume} {128}},\ \bibinfo {pages} {7043}
  (\bibinfo {year} {2006})}\BibitemShut {NoStop}%
\bibitem [{\citenamefont {K{\"o}ksal}\ and\ \citenamefont
  {Ko{\c{c}}}(2016)}]{koksal2016optical}%
  \BibitemOpen
  \bibfield  {author} {\bibinfo {author} {\bibfnamefont {K.}~\bibnamefont
  {K{\"o}ksal}}\ and\ \bibinfo {author} {\bibfnamefont {F.}~\bibnamefont
  {Ko{\c{c}}}},\ }\href@noop {} {\bibfield  {journal} {\bibinfo  {journal}
  {Philosophical Magazine}\ }\textbf {\bibinfo {volume} {96}},\ \bibinfo
  {pages} {2686} (\bibinfo {year} {2016})}\BibitemShut {NoStop}%
\bibitem [{\citenamefont {K{\"o}ksal}\ and\ \citenamefont
  {Ko{\c{c}}}(2017{\natexlab{a}})}]{koksal2017effect}%
  \BibitemOpen
  \bibfield  {author} {\bibinfo {author} {\bibfnamefont {K.}~\bibnamefont
  {K{\"o}ksal}}\ and\ \bibinfo {author} {\bibfnamefont {F.}~\bibnamefont
  {Ko{\c{c}}}},\ }\href@noop {} {\bibfield  {journal} {\bibinfo  {journal}
  {Computational and Theoretical Chemistry}\ }\textbf {\bibinfo {volume}
  {1099}},\ \bibinfo {pages} {203} (\bibinfo {year}
  {2017}{\natexlab{a}})}\BibitemShut {NoStop}%
\bibitem [{\citenamefont {K{\"o}ksal}\ and\ \citenamefont
  {Ko{\c{c}}}(2017{\natexlab{b}})}]{koksal2017spin}%
  \BibitemOpen
  \bibfield  {author} {\bibinfo {author} {\bibfnamefont {K.}~\bibnamefont
  {K{\"o}ksal}}\ and\ \bibinfo {author} {\bibfnamefont {F.}~\bibnamefont
  {Ko{\c{c}}}},\ }\href@noop {} {\bibfield  {journal} {\bibinfo  {journal}
  {Computational and Theoretical Chemistry}\ }\textbf {\bibinfo {volume}
  {1105}},\ \bibinfo {pages} {27} (\bibinfo {year}
  {2017}{\natexlab{b}})}\BibitemShut {NoStop}%
\bibitem [{\citenamefont {Ko{\c{c}}}\ and\ \citenamefont
  {K{\"o}ksal}(2015)}]{kocc2015quantum}%
  \BibitemOpen
  \bibfield  {author} {\bibinfo {author} {\bibfnamefont {F.}~\bibnamefont
  {Ko{\c{c}}}}\ and\ \bibinfo {author} {\bibfnamefont {K.}~\bibnamefont
  {K{\"o}ksal}},\ }\href@noop {} {\bibfield  {journal} {\bibinfo  {journal}
  {Superlattices and Microstructures}\ }\textbf {\bibinfo {volume} {85}},\
  \bibinfo {pages} {599} (\bibinfo {year} {2015})}\BibitemShut {NoStop}%
\bibitem [{\citenamefont {Ko{\c{c}}}\ and\ \citenamefont
  {K{\"o}ksal}(2017)}]{kocc2017mapping}%
  \BibitemOpen
  \bibfield  {author} {\bibinfo {author} {\bibfnamefont {F.}~\bibnamefont
  {Ko{\c{c}}}}\ and\ \bibinfo {author} {\bibfnamefont {K.}~\bibnamefont
  {K{\"o}ksal}},\ }\href@noop {} {\bibfield  {journal} {\bibinfo  {journal}
  {Computational and Theoretical Chemistry}\ }\textbf {\bibinfo {volume}
  {1117}},\ \bibinfo {pages} {87} (\bibinfo {year} {2017})}\BibitemShut
  {NoStop}%
\bibitem [{\citenamefont {Chulkov}\ \emph {et~al.}(1997)\citenamefont
  {Chulkov}, \citenamefont {Silkin},\ and\ \citenamefont
  {Echenique}}]{chulkov1997image}%
  \BibitemOpen
  \bibfield  {author} {\bibinfo {author} {\bibfnamefont {E.}~\bibnamefont
  {Chulkov}}, \bibinfo {author} {\bibfnamefont {V.}~\bibnamefont {Silkin}}, \
  and\ \bibinfo {author} {\bibfnamefont {P.}~\bibnamefont {Echenique}},\
  }\href@noop {} {\bibfield  {journal} {\bibinfo  {journal} {Surface science}\
  }\textbf {\bibinfo {volume} {391}},\ \bibinfo {pages} {L1217} (\bibinfo
  {year} {1997})}\BibitemShut {NoStop}%
\bibitem [{\citenamefont {Chulkov}\ \emph {et~al.}(1999)\citenamefont
  {Chulkov}, \citenamefont {Silkin},\ and\ \citenamefont
  {Echenique}}]{chulkov1999image}%
  \BibitemOpen
  \bibfield  {author} {\bibinfo {author} {\bibfnamefont {E.}~\bibnamefont
  {Chulkov}}, \bibinfo {author} {\bibfnamefont {V.}~\bibnamefont {Silkin}}, \
  and\ \bibinfo {author} {\bibfnamefont {P.}~\bibnamefont {Echenique}},\
  }\href@noop {} {\bibfield  {journal} {\bibinfo  {journal} {Surface science}\
  }\textbf {\bibinfo {volume} {437}},\ \bibinfo {pages} {330} (\bibinfo {year}
  {1999})}\BibitemShut {NoStop}%
\bibitem [{\citenamefont {K{\"o}ksal}\ and\ \citenamefont
  {Berakdar}(2012)}]{koksal2012charge}%
  \BibitemOpen
  \bibfield  {author} {\bibinfo {author} {\bibfnamefont {K.}~\bibnamefont
  {K{\"o}ksal}}\ and\ \bibinfo {author} {\bibfnamefont {J.}~\bibnamefont
  {Berakdar}},\ }\href@noop {} {\bibfield  {journal} {\bibinfo  {journal}
  {Physical Review A}\ }\textbf {\bibinfo {volume} {86}},\ \bibinfo {pages}
  {063812} (\bibinfo {year} {2012})}\BibitemShut {NoStop}%
\end{thebibliography}%

\end{document}